\documentstyle[12pt,epsfig]{article}
\def\lapproxeq{\lower .7ex\hbox{$\;\stackrel{\textstyle
<}{\sim}\;$}}
\def\gapproxeq{\lower .7ex\hbox{$\;\stackrel{\textstyle
>}{\sim}\;$}}
\begin{document}
%
%\begin{titlepage}
%\pagestyle{empty}
%\vspace*{2cm}
\begin{center}
{\LARGE\bf  Spin dependent structure function $g_1$ at
low $x$  and  low $Q^2$}
\vspace{1.1cm}\\

%         {\sc B.~Bade\l{}ek $^a$} and {\sc J.~Kwieci\'nski $^b$} \\
%%         {\sc J.~Kwieci\'nski $^b$},\\
%\vspace{0.3cm}
%$^a$ {\it Department of Physics, Uppsala University, P.O.Box 530,
%751 21 Uppsala, Sweden} \\
%%\noindent
%%\hspace{2mm}
%{\it and Institute of Experimental Physics, Warsaw University, Ho\.za 69,
%00-681 Warsaw, Poland (e-mail: badelek@tsl.uu.se)}\\

%$^b$ {\it Department of Theoretical Physics, 
%H.~Niewodnicza\'nski Institute of Nuclear Physics, \\
%Radzikowskiego 152, 31-342 Cracow, Poland (e-mail:
%jkwiecin@solaris.ifj.edu.pl)} \\

{\sc B.~Bade\l{}ek $^{a,b}$} {\sc J. Kiryluk $^b$}
and {\sc J.~Kwieci\'nski $^c$} \\
\end{center}

\vspace{0.3cm}
\noindent
$^a$ {\it Department of Physics, Uppsala University, P.O.Box 530,
751 21 Uppsala, Sweden} \\
\noindent
$^b$ {\it Institute of Experimental Physics, Warsaw University, Ho\.za 69,
00-681 Warsaw,\\ ~~~~ Poland}\\
\noindent
$^c$ {\it Department of Theoretical Physics,
H.~Niewodnicza\'nski Institute of Nuclear Physics,
Radzikowskiego 152, 31-342 Cracow, Poland} \\
\begin{center}
\vskip-3mm
{ e-mail addresses: badelek@fuw.edu.pl, kiryluk@fuw.edu.pl, 
jkwiecin@solaris.ifj.edu.pl}

\end{center}
\vskip5mm
\begin{abstract}
Theoretical description of the spin dependent structure function 
$g_1(x,Q^2)$ in the region of low values of $x$ and $Q^2$ is presented. 
It contains the Vector Meson Dominance 
contribution and the QCD improved parton model suitably extended 
to the low $Q^2$ domain.  Theoretical 
predictions are compared with the recent experimental data in the low $x$, 
low $Q^2$ region. 
\end{abstract}
%\vskip5mm

\section*{1. Introduction}
\label{section:intro}

\noindent
Measurements of polarised deep inelastic lepton--nucleon scattering
have determined the cross 
section asymmetries $A_1$ and spin dependent structure functions $g_1$
of the proton, deuteron and neutron
in a wide kinematic range of $Q^2$ and $x$.\footnote{Here, as usual, 
$x=Q^2/(2pq)$ where $Q^2=-q^2$ with $q$ and $p$ 
denoting the four momentum transfer between leptons and the four momentum 
of the nucleon respectively.} This allowed a verification 
of sum rules, like e.g. the Bjorken sum rule which is a fundamental relation
in the QCD, and the Ellis--Jaffe sum rules. Evaluation of the sum rules 
requires knowledge of the structure functions $g_1$ over the entire 
region of $x$
as well as their evolution to a common value of $Q^2$.
Since the experimentally accessible $x$ range is limited, extrapolations
to $x=$ 0 and $x=$ 1 are necessary. Of these the former is critical since 
the small $x$ behaviour of $g_1(x)$ is theoretically not well established
and the relevant contribution to the sum rules' integral may in principle 
be large. \\
%However it is better to have theoretical predictions for the $g_1$ 
%over a full range of $x$ than for its one moment only. 
%Especially the knowledge of the structure function $g_1$ at low $x$, 
%i.e. at high parton densities, is very interesting in itself. 
%The new dynamical mechanisms may be revealed, as for
%the unpolarised case where they are being intensively studied \cite{hera}.

Theoretical predictions for the structure function $g_1$ over a full 
range of $x$ 
are even more interesting than for its first moment, 
especially at low $x$, i.e. at high parton densities, where the new dynamical 
mechanisms may be revealed. 
Theoretical and experimental studies at low $x$ in the polarised case
are thus awaited for. A possible future polarising the proton beam at 
HERA would 
be a milestone in this field. \\

In the fixed target experiments the low values of $x$ are reached 
by lowering at the same time the values of $Q^2$.  Theoretical
 analysis of these data therefore requires a suitable extrapolation of the 
 structure function to the low $Q^2$ region. 
Low $Q^2$ phenomena and in particular a
transition from the perturbative (large $Q^2$) to the nonperturbative 
(low $Q^2$, including $Q^2$=0) region is actively investigated
in the spin insensitive experiments. In spite of a wealth of data 
and of a wide spectrum of ideas this
field is still a major challenge in high energy physics \cite{hera}. 
Among the spin sensitive experiments the only available low $Q^2$ data are from
the E143 experiment at SLAC \cite{e143} (moderate $x$ and low $Q^2$) 
and now also from the SMC at CERN \cite{optimal,t15} (low $x$ and low $Q^2$).
 In the low $Q^2$ region one can expect 
 that dynamical mechanisms, like the Vector Meson Dominance (VMD), 
 can play an important role. 
For large $Q^2$ the VMD contribution to $g_1$ gives a power correction 
 term and can usually be neglected.  
%It is however expected to play important role in the low $Q^2$ region.  
Moreover, the partonic contribution to $g_1$ 
 which controls the structure functions in the deep inelastic domain has to be 
 suitably extended in order to get its extrapolation to the low $Q^2$ region.  
The latter component 
  will be expressed  in terms of the unintegrated (spin dependent) parton 
  distributions and we show that the corresponding representation of $g_1$ 
 can be easily extrapolated to the low $Q^2$ region. 
The main purpose of our paper is therefore to construct 
 the structure function $g_1(x,Q^2)$ which would include the VMD
 and the (QCD improved) parton model contributions.  \\

The content of the paper is as follows: in the next Section
we present the data on $g_1$ and  
 comment on the Regge model predictions for $g_1$
 which are often being used for
$x=$ 0 extrapolations. In Sec.3 we briefly present a formalism describing 
$g_1$ in terms of the 
unintegrated spin dependent
parton distributions, incorporating the leading order Altarelli--Parisi 
evolution and the double logarithmic ln$^2(1/x)$ resummation at low $x$.
In Sec.4 we discuss the Vector Meson Dominance part of the $g_1$
which has to be included in that region since, 
as it has already been pointed out above,  
for the fixed target experiments low values of $x$ are correlated with 
the low values of $Q^2$. 
Numerical results are also presented there. 
Finally, in Sec.5 we give a summary of our analysis. 
\\

\section*{2. The $g_1$ data}
\label{section:data}

Several experiments contributed to the spin structure function $g_1$ 
measurements on different targets and over different kinematic intervals. 
As a result, for proton and deuteron, $g_1$ was measured for 
0.00006 $< x <$ 0.8 by the EMC \cite{emc}, SMC \cite{optimal,t15}, 
E143 \cite{e143}, E155 \cite{e155} 
and HERMES \cite{hermesp}.  For neutron, $g_1$ was measured for 
0.014 $< x < $ 0.8 by the E142 \cite{e142}, E154 \cite{e154} and 
HERMES \cite{hermesn}. 
A summary of $xg_1^{p,d}(x)$ data at the measured $Q^2$ values 
is presented in Fig.1. 
%Only statistical errors are marked.
%A high statistical accuracy was obtained in 
%the kinematic range of the SLAC and DESY experiments, i.e. for 
%$x \gapproxeq$ 0.01. \\

\begin{figure}[ht]
\epsfxsize=8cm
\hfil
\epsffile{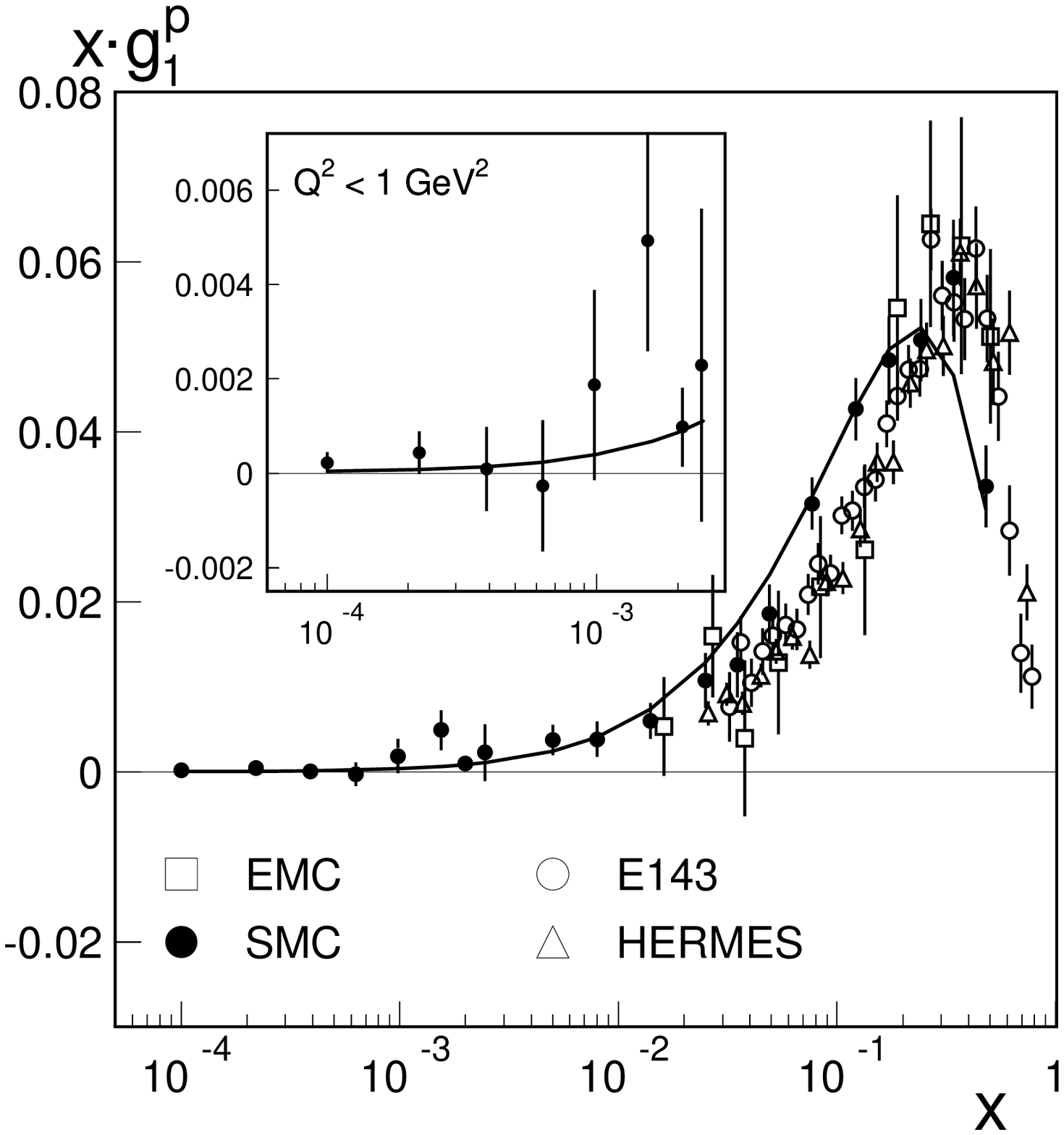}
\epsfxsize=8cm
\epsffile{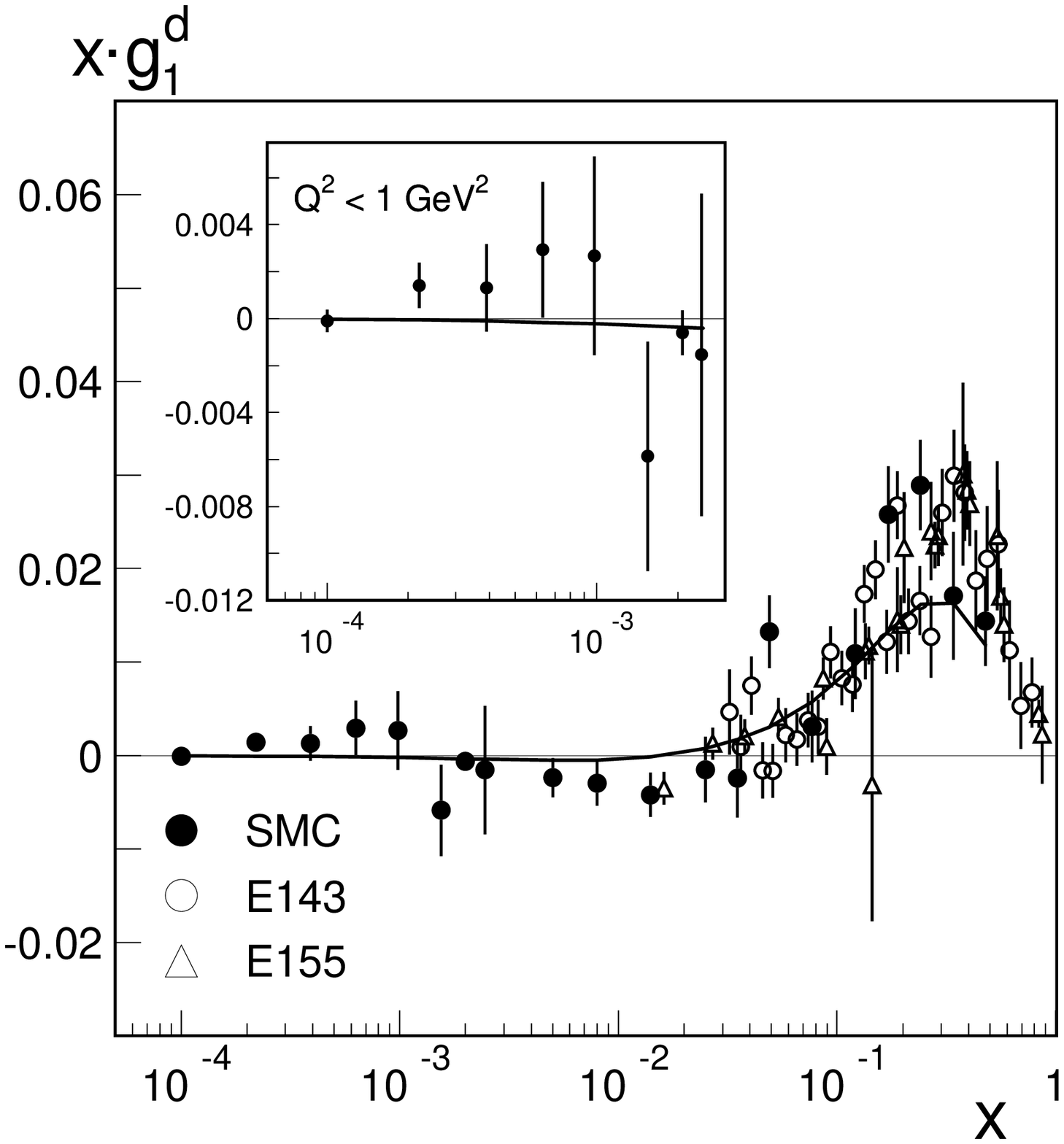}
\hfil
\caption{Summary of the $xg_1$ measurements for the proton and for the 
deuteron as a function of $x$ at the measured $Q^2$ obtained with different
experiments. The inserted figures show the SMC data for which 
$Q^2 < $ 1 GeV$^2$.  Errors are statistical. The curves, 
calculated at $x$ and $Q^2$ values at the SMC measurements result from the 
model described in Sec.3.
\label{fig:xg1pd}}
\end{figure}

%Typically for fixed target experiments $x$ is strongly correlated with $Q^2$. 
\noindent
For the SMC data, $\langle x \rangle$ = 0.0001 corresponds to 
$\langle Q^2 \rangle$ = 0.02 GeV$^2$. 
In other expe\-ri\-ments $g_1$ was measured with high statistical
accuracy for $x\gapproxeq$0.01 and for $Q^2 >$ 1 GeV$^2$ only.\footnote
{The E143 measured the asymmetry $A_1^{p,d,n}$ for 0.024 $< x <$ 0.205
 and 0.31 $< Q^2 <$ 1 GeV$^2$ but $g_1$ was not extracted from those data.} 
\noindent
 We do not present $g_1^n$ as  there are no direct measurements 
for $x < 0.01$ i.e. in the low $x$ region.  \\

 The lowest $x$ and $Q^2$ 
region was explored by the SMC due to a high energy of the 
muon beam and implementation of a dedicated low $x$ trigger.  
The results of the SMC presented in Fig.1 come from two different 
analyses \cite{optimal,t15} which join at $x\sim$0.002. 
%In the standard SMC analysis \cite{optimal} the kinematic range was 
%0.0008 $< x <$ 0.7. Events with lower values of $x$ were not measured to 
%avoid a contamination with elastic muon scattering 
%off atomic electrons which is a dominant process at $x$=0.000545. 
%To overcome this difficulty a new trigger was built in which both a 
%signal from a hadronic
%part of the calorimeter and a scattered muon were demanded \cite{t15}.  
%This together with off-line selections removed most of $~\mu e$ events
%and considerably reduced the radiative background. 
%As a result the SMC kinematic region extends down to $x$ = 0.00006. 
It should be noted that a direct result of the measurements is the
virtual photon--nucleon asymmetry, $A_1$.
To get the $g_1$ one has to use the relation 
$g_1 = A_1\cdot F_1 \equiv A_1\cdot F_2/[2x(1+R)]$, where $F_2=F_L + F_T$, 
$R=F_L/F_T$ and $F_T=2xF_1$ with $F_L$ and $F_T$ denoting the unpolarised 
nucleon structure functions corresponding  to longitudinal and transverse 
polarisations of the virtual photon respectively. 
Unfortunately there have been no direct measurements of  $F_2$ and $R$
in the kinematic region of the low $x$ and low $Q^2$ SMC data, i.e. for 
$0.00006< x < 0.003$ and $0.01 < Q^2 < 1$ GeV$^2$.
Thus the SMC used the model \cite{BBJK2} for the $F_2$ and a parametrisation 
of Ref.\cite{r} for $R$ so their results for $g_1$ are model--dependent.  \\
%The inserts in Fig.1 show the SMC data for which $Q^2 < $1 GeV$^2$. \\

 The new low $x$ data of the SMC include the kinematic region where  
$W^2=(p+q)^2$ is high, $W^2\gapproxeq$ 100 GeV$^2$ and much larger than $Q^2$. 
Thus one should expect that the Regge model should be applicable there. 
According to the Regge model, $g_1(x,Q^2) \sim x^{-\alpha}$ for 
$x \rightarrow 0$ and fixed $Q^2$, where 
$\alpha$  denotes the intercept of the Regge pole trajectory corresponding 
to axial vector mesons.  It is expected that $\alpha \sim 0$ for both 
$I=0$ and $I=1$ trajectories, \cite{hei}. 
This behaviour of $g_1$ should go smoothly to the $W^{2 \alpha}$  
dependence for $Q^2 \rightarrow 0$. 
Other considerations related to the Regge theory
predict  $g_1\sim${ln}$x$, \cite{clo_rob}, 
while the model based on exchange of two nonperturbative gluons gives 
$g_1\sim$ 2~{ln}(1/$x$)--1, \cite{bass_land}. A perverse behaviour,
$g_1\sim$1/$(x$ln$^2x)$, recalled in \cite{clo_rob}, is not valid  
for $g_1$, \cite{misha}.\\

%However a low intensity of the beam resulted in a rather poor statistical accuracy in comparison with experiments with an electron beam. 
% At the lowest $x$, $<Q^2>=$0.02 GeV$^2$.
% The E155 results for the deuteron were obtained at three
%different values of the electron scattering angle, corresponding
%to different values of $Q^2$ at fixed $x$. 
%\section*{3. Regge model predictions}
%\label{section:regge}
%This model gives the $x$ (or $W^2$) dependence of $g_1$ at fixed $Q^2$.

In the kinematic range of the SMC data $W^2$ changes very little:
from about 100 GeV$^2$ at $x=$ 0.1 to about 220 GeV$^2$ at $x=$ 0.0001,
contrary to a quite strong change of $Q^2$ (from about 20 GeV$^2$ 
to about 0.01 GeV$^2$ respectively).  
This means that the new SMC measurements cannot test the Regge 
behaviour of $g_1$ through the $x$ dependence of the latter, without 
additional assumptions about the $Q^2$ dependence of $g_1$. 
A model which allows extrapolation of $g_1$ to the 
low $Q^2$ region 
is described in the next Section.

%For completeness we shall only list the Regge predictions. 

\section*{3. Partonic contribution to $g_1$}
\label{section:qcd}

\noindent 
In the region of large values of $Q^2$ the spin depedent structure functions 
are described by the QCD improved parton model \cite{reya}.   
In this model $g_1 \equiv g_1^{part}$, where $g_1^{part}$ is related 
in a standard
way to the polarised quark and antiquark distributions $\Delta q_i$ 
and $\Delta \bar q_i$ corresponding to the quark (antiquark) 
flavour $i$:  
\begin{equation}
g_1^{part}(x,Q^2) = {1\over 2}\sum_{i=u,d,s} e_i^2\left[\Delta q_i(x,Q^2) + 
\Delta \bar q_i(x,Q^2)
\right].
\label{gp1}
\end{equation}
%where $\Delta u_v(x,Q^2) = \Delta p_{u_v}(x,Q^2)$ etc.
In what follows we assume $\Delta \bar q_{u} =\Delta \bar q_{d}$ and  set the  
number of flavours equal 3. \\

In perturbative QCD the structure function $g_1^{part}$
is controlled at low $x$ by the {\it double} logarithmic
ln$^2(1/x)$ contributions  
i.e. by those terms of the perturbative expansion which cor\-res\-pond to
the powers of ln$^2(1/x)$ at each order of the expansion
\cite{BARTSNS}.  
%As has recently been pointed out, \cite{pioneers},
%small $x$ behaviour of
%the spin dependent structure function $g_1(x,Q^2)$, is
%controlled by the double logarithmic terms, i.e. by those
%terms which correspond to powers of $\alpha_s${ln} $(1/x)$.
%Such terms also appear in the nonsinglet part of the
%spin independent structure function,
%$F_2^{ns}(x,Q^2)$, but a leading small $x$ behaviour of it
%which those terms generate is overridden by the
%contribution of the relevant Regge pole.
%In case of $g_1$, the Regge poles are expected
%to have a low intercept and thus cannot mask a more singular
%{ln}$(1/x)$ behaviour.   
%The data presented here offer
%a unique possibility of testing these special effects.
%The $g_1^p(x,Q^2)$
%has thus been calculated using a unified scheme,
%incorporating the complete leading order
%Altarelli--Parisi evolution at finite $x$ and double logarthmic
%ln$^2$(1/$x$) effects at $x\rightarrow$0, \cite{bb_jk_bz}.
%The resulting $g_1^p$ had the right behaviour at $Q^2\rightarrow$0
%and could thus be extrapolated to the region of low $Q^2$.
%However, nonperturbative
%contributions to $g_1^p$ may also be important in
%the $Q^2$=0 limit. They are being estimated, \cite{ht}. 
It is convenient to discuss the ln$^2(1/x)$ resummation
using the formalism of the   unintegrated (spin dependent) parton
distributions
$f_j(x^{\prime},k^2)$ ($j=u_{v},d_{v},\bar u,\bar d,\bar s,g$) where
$k^2$
is the transverse momentum squared of the parton $j$ and $x^{\prime}$
the
longitudinal momentum fraction of the parent nucleon carried by a parton
\cite{BBJK,JKAPP,BZIAJA}.
The conventional (integrated) distributions $\Delta p_j(x,Q^2)$ (i.e. 
$\Delta q_u=\Delta p_{u_v} + \Delta p_{\bar u}, ~
\Delta \bar q_u = \Delta p_{\bar u}$ etc. for quarks, antiquarks 
and gluons) are 
related in the
following way  to the unintegrated distributions $f_j(x^{\prime},k^2)$:
\begin{equation}
 \Delta p_j(x,Q^2)=\Delta p_j^{0}(x)+
 \int_{k_0^2}^{W^2}{dk^2\over k^2}f_j(x^{\prime}=
x(1+{k^2\over Q^2}),k^2)
\label{dpi}
\end{equation}
Here $\Delta p_j^{0}(x)$ denote the
nonperturbative parts of the of the distributions, corresponding to
$k^2 < k^2_0$ and the parameter $k_0^2$ is the infrared cut-off
($k_0^2 \sim $1 GeV$^2$). They are treated semiphenomenologically and 
parametrised in the form used in 
Refs \cite{BBJK,JKAPP,BZIAJA}:

\begin{equation}
\Delta p_j^0(x) = C_j (1-x)^{\eta_j}
\label{input}
\end{equation}   \\
In Eq.(\ref{input}) we
%assuming the parametrisations of the 
%input distributions $\Delta p_i^0(x)$ as in Eq. (\ref{input}). 
assumed ${\eta_u}_v = {\eta_d}_v = $3, $\eta_{\bar u} = \eta_{\bar s} =$ 7 
and
$\eta_g =$ 5. We also used $k_0^2 = $1 GeV$^2$. The normalisation constants $C_j$ were determined by imposing
the Bjorken sum rule for $\Delta u_v^{0} - \Delta d_v^{0}$ and
by requiring that the first moments of all other distributions are the same
as those determined from the QCD analysis of \cite{stratmann}.  \\
 
%The structure function $g_1(x,Q^2)$ is related 
%in the following way to $f(x^{\prime},k^2)$:
%
%\begin{equation}
%g_1(x,Q^2)=g_1^{(0)}(x) + calka ... bla bla bla 
%\end{equation}
%where $g_1^{(0)}(x)$ is the nonperturbative part of the structure function. \\ 

The unintegrated distributions $f_j(x^{\prime},k^2)$ are the solutions of the 
integral equations \cite{BBJK,JKAPP,BZIAJA} which embody both the LO Altarelli-Parisi 
 evolution \cite{AP} and the double 
ln$^2(1/x^{\prime})$ resummation at small $x^{\prime}$. 
%The sum of ln$^2(1/x)$
%terms is  generated by the corresponding integral equations
%for the functions $f_i(x^{\prime},k^2)$ \cite{BBJK,JKAPP,BZIAJA,MANAR}. 
These equations combined with equations (\ref{gp1}) and (\ref{dpi}) 
lead to approximate $x^{-\lambda}$ behaviour of the $g_1^{part}$ in 
the $x \rightarrow 
0$ limit, 
with $\lambda \sim 0.3$ and 
$\lambda \sim 1$ for the nonsinglet and singlet parts 
respectively which is more singular at low $x$ than 
that   generated by the (nonperturbative)  Regge pole exchanges 
\footnote{ To be precise the singular $x^{-\lambda}$ 
behaviour with $\lambda \sim 1$ for singlet and gluon spin dependent 
distributions does hold in the approximation when only the 
ladder diagrams are retained \cite{JKAPP}. Complete double logarithmic 
$ln^2(1/x)$  resummation which 
includes also the non-ladder bremmstrahlung diagrams generates less singular 
behaviour of these distributions \cite{BZIAJA}.}.  
The double ln$^2(1/x)$ effects are presumably not important
in the $W^2$ range of the fixed target experiments (cf. Fig.2 in
\cite{BBJK} and Fig. 6 in \cite{BZIAJA})
 but they significantly affect $g_1$ in the low $x$ region 
which may be probed at the polarised HERA, \cite{BBJK,JKAPP,BZIAJA}.
However the formalism based on the 
unintegrated distributions employed here is very suitable 
for extrapolating 
$g_1$ to the region of low $Q^2$ at fixed $W^2$ \cite{BBJK}
.  \\

Formulae (\ref{gp1}) and (\ref{dpi}) define partonic contribution 
to the 
structure function $g_1(x,Q^2)$.
% which, as it was  shown above can be 
%automatically extrapolated to the region of low $Q^2$ for 
%fixed and large $W^2$. \\
Since $x(1+k^2/Q^2) \rightarrow k^2/W^2$ for $Q^2 \rightarrow 0$  
in the integrand in Eq. (\ref{dpi}) and since $k^2 >k_0^2$ there,  
the $g_1^{part}(x,Q^2)$ defined by Eqs (\ref{gp1}) and (\ref{dpi})
can be smoothly extrapolated to 
the low $Q^2$ region, including $Q^2=0$. In that limit, the 
$g_1$ should be a finite function of $W^2$, free from any
kinematical singularities or zeros. The extrapolation,
valid for fixed and large $W^2$, can thus be done  for the $g_1^{part}(x,Q^2)$
given by Eqs (\ref{gp1}) and (\ref{dpi}) provided that
nonperturbative parts of the parton distributions 
$\Delta p_j^{0}(x)$ are free from
kinematical singularities at $x=0$, as in the parametrisations 
defined by Eq. (\ref{input}). If $\Delta p_j^{0}(x)$ contain kinematical singularities
at $x=0$ then
% one may just use the
%same prescription as that which was adopted in ref.\cite{BBJK2}, i.e.
one may replace  $\Delta p_j^{0}(x)$
with $\Delta p_j^{0}(\bar x)$ where
$\bar x = x\left(1+{k_0^2/ Q^2}\right)$
and leave remaining parts of the calculation unchanged.
After this simple rearrangement the structure function $g_1^{part}(x,Q^2)$
 can be
extrapolated to the low $Q^2$ region (for fixed $W^2$) including 
the point $Q^2=0$.
Possibility of extrapolation to $Q^2=0$ is an important
property of the formalism based on the unintegrated parton distributions. \\

We solved equations for the functions $f_i(x^{\prime},k^2)$ 
\cite{BBJK,JKAPP,BZIAJA} and calculated the $g_1^{part}(x,Q^2)$ from Eqs
(\ref{gp1}) and (\ref{dpi}) using the parametrisation (\ref{input}). To be precise 
we solved equations which resummed only the ladder diagrams contributions 
in that part which corresponded to the double ln$^2(1/x)$ resummation but 
this approximmation was completely adequate for the values of $W^2$ which are 
relevant for the fixed target experiments.  Let us also remind that 
equations for the functions $f_i(x,Q^2)$ \cite{BBJK,JKAPP,BZIAJA} combined with equations (\ref{gp1},\ref{dpi}) 
are a generalisation of the LO QCD evolution equations \cite{AP} for polarised parton 
densities and for moderately small and large values of $x$ are equivalent 
to these equations.  \\

As a consequence $g_1^{part}$ calculated at $x$ and $Q^2$ values of the
SMC measurement gives a reasonable description of the SMC data on  
$g_1^{p,d}(x,Q^2)$, cf. Fig.1 (it does not reproduce at the same time other measurements 
equally well due to differences in $Q^2$ values between the experiments). 
For the sake of the comparison the calculated $g_1^{part}$
was extrapolated to low values of $Q^2$ since all the data with 
$x\lapproxeq$0.001 have
$Q^2 <$~1 GeV$^2$. However the (extrapolated) $g_1^{part}$ may not be 
the only contribution to $g_1$ in the low $Q^2$ domain.

%\section*{5. Nonperturbative contributions to $g_1$ at low $Q^2$}
\section*{4. Vector Meson Dominance contribution to $g_1$ }
\label{section:vmd}

One expects that in the low $Q^2$ region an important role may be played 
by the VMD mechanism. The structure function should thus be represented
by the sum of the partonic and VMD contributions, i.e.
\begin{equation}
g_1(x,Q^2)=g_{1}^{VMD}(x,Q^2) + g_{1}^{part}(x,Q^2)
\label{g1tot}
\end{equation}

\noindent
The VMD contribution 
%$g_{1}^{VMD}(x,Q^2)$  
to $g_1(x,Q^2)$ can be written as: 
\begin{equation}
g_{1}^{VMD}(x,Q^2)= {pq\over 4 \pi} \sum_{v=\rho,\omega,\phi} 
{ m_v^4 \Delta \sigma_{v}(W^2) \over \gamma_v^2 
(Q^2+m_v^2)^2} 
\label{vmdg1}
\end{equation}
In this formula the constants 
$\gamma_v^2$ are determined from the leptonic widths of the vector mesons 
\cite{BAUER} and $m_v$ denotes the mass of the vector meson $v$.  
The  cross 
sections $\Delta \sigma_{v}(W^2)$ are for high energy $W^2$ given as the 
following combinations of the spin dependent total  cross sections:
\begin{equation}
\Delta \sigma_{v} = {{\sigma_{1/2} - \sigma_{3/2} }\over 2}
\label{delsigma}
\end{equation}
where $\sigma_{1/2}$ and $\sigma_{3/2}$ correspond to the total  
vector meson - nucleon cross
 sections with the projections of the total spin on the vector meson 
momentum equal 1/2 and 3/2 respectively \cite{IOFFE}. 
Unfortunately the cross-sections $\Delta \sigma_{v}$ are unknown.    
\noindent
  In order to estimate the VMD contribution, 
$g_{1}^{{VMD}}(x,Q^2)$,
we assume that the cross sections $\Delta \sigma _{v}$ are proportional 
to the appropriate combinations of the nonperturbative contributions  
$\Delta p_j^0(x)$, defined by Eq.(\ref{input}), to the polarised quark 
and antiquark distributions. For the proton we assume: 
$$
{pq\over 4 \pi} \sum_{v=\rho,\omega}{m_v^4 \Delta \sigma_{v} 
\over \gamma_v^2 (Q^2 + m_v^2)^2} = 
$$

%\begin{equation}
%C\left[{4\over 9} \left(\Delta u^0(x) + \Delta \bar u^{0}(x)\right) 
%+{1\over 9} \left(\Delta d^{0}(x) + \Delta \bar d^0(x)\right)\right]
%{m_{\rho}^4\over (Q^2+m_{\rho}^2)^2}
%\label{nsvmd}
%\end{equation}
%\begin{equation}
%{pq\over 4 \pi} {m_{\phi}^4 \Delta \sigma_{\phi p} \over \gamma_{\phi}^2 
% (Q^2 + m_{\rho}^2)^2} = 
%C{2\over 9}(\Delta s^0(x) + \Delta \bar s^0(x)){m_{\phi}^4 \over (Q^2 +m_{\phi}^2)^2} 
%\end{equation}

\begin{equation}
C \left  [ {4\over 9} \left(\Delta u^0_v(x) + 2\Delta \bar u^{0}(x)\right) 
+{1\over 9} \left(\Delta d^{0}_v(x) + 2\Delta \bar u^0(x)\right)\right]
{m_{\rho}^4\over (Q^2+m_{\rho}^2)^2}
\label{nsvmd}
\end{equation}

\begin{equation}
{pq\over 4 \pi} {m_{\phi}^4 \Delta \sigma_{\phi p} \over \gamma_{\phi}^2 
 (Q^2 + m_{\rho}^2)^2} = 
C{2\over 9}\Delta \bar s^0(x){m_{\phi}^4 \over (Q^2 +m_{\phi}^2)^2} 
\label{svmd}
\end{equation}    

\noindent 
%$p$ and $q$ denote the four-momenta of the proton and virtual photon
%respectively
where $\Delta u^0(x)=\Delta p_u^{0}(x)$ etc. All distributions 
are parton distributions in the proton.     
%The parametrisation defining the nonpertur   
%The parametrisation defining the nonperturbative 
%parton distributions $\Delta p_i^0(x)$ is of the form used in 
%refs \cite{BBJK,BZIAJA}:
%
%\begin{equation}
%\Delta p_i^0(x) = C_i (1-x)^{\eta_i}
%\label{input}
%\end{equation}
% 
%as in eq. (\ref{input}) 
%\noindent
%It should be noted (cf. equations (\ref{nsvmd},\ref{svmd},\ref{input}) that 
The distributions $\Delta p_j^0(x)$, Eq. (\ref{input}), behave as $x^0$ for 
$x\rightarrow 0$. As a result
 the  cross sections $\Delta \sigma_{v}$ behave as 
%$\Delta \sigma_{v}$ would behave  as 
$1/W^2$ at large $W^2$ that corresponds 
to the assumption 
that  the corresponding Regge trajectories have their intercepts equal
to zero. 
We include exact $x$ dependence of the nonperturbative 
(spin dependent) parton distributions $\Delta p_j^0(x)$ and not only their (constant) 
$x \rightarrow 0$ limits, $C_j$.   This gives  an 
extension of the VMD model to the region of moderately small values of $x$.  
Formally this  means that we allow additional $Q^2$ dependence of the 
cross-sections $\Delta \sigma_{v}$ in terms which are non-leading in the large 
$W^2$ limit, i.e.  vanish faster than 
 $1/W^2$. 
 \\

  We shall vary the parameter $C$ 
in Eqs (\ref{nsvmd}) and (\ref{svmd}) and analyse a dependence 
of the structure function $g_1(x,Q^2)$ upon the value of this 
parameter. 
It should be noted that the VMD part of $g_1$ vanishes at large $Q^2$ 
as $1/Q^4$ (contrary to the $g_1^{part}$ 
which scales {\it modulo} logarithmic corrections) but 
it may be a dominant contribution at (very) low $Q^2$ as it is the case 
for the unpolarised structure functions. 
For low $Q^2$ we expect a dominance of the VMD part of $g_1$.
In analogy with the unpolarised case we expect that it should exhaust 
about 80 $\%$ of the total $g_1$.   \\

%path to fig in Warsaw: ~kiryluk/g1_model/paper/a1p.smc.eps 
%                       ~kiryluk/g1_model/paper/a1p.16.2.eps

\begin{figure}[ht]
\epsfxsize=8cm
\hfil
\epsffile{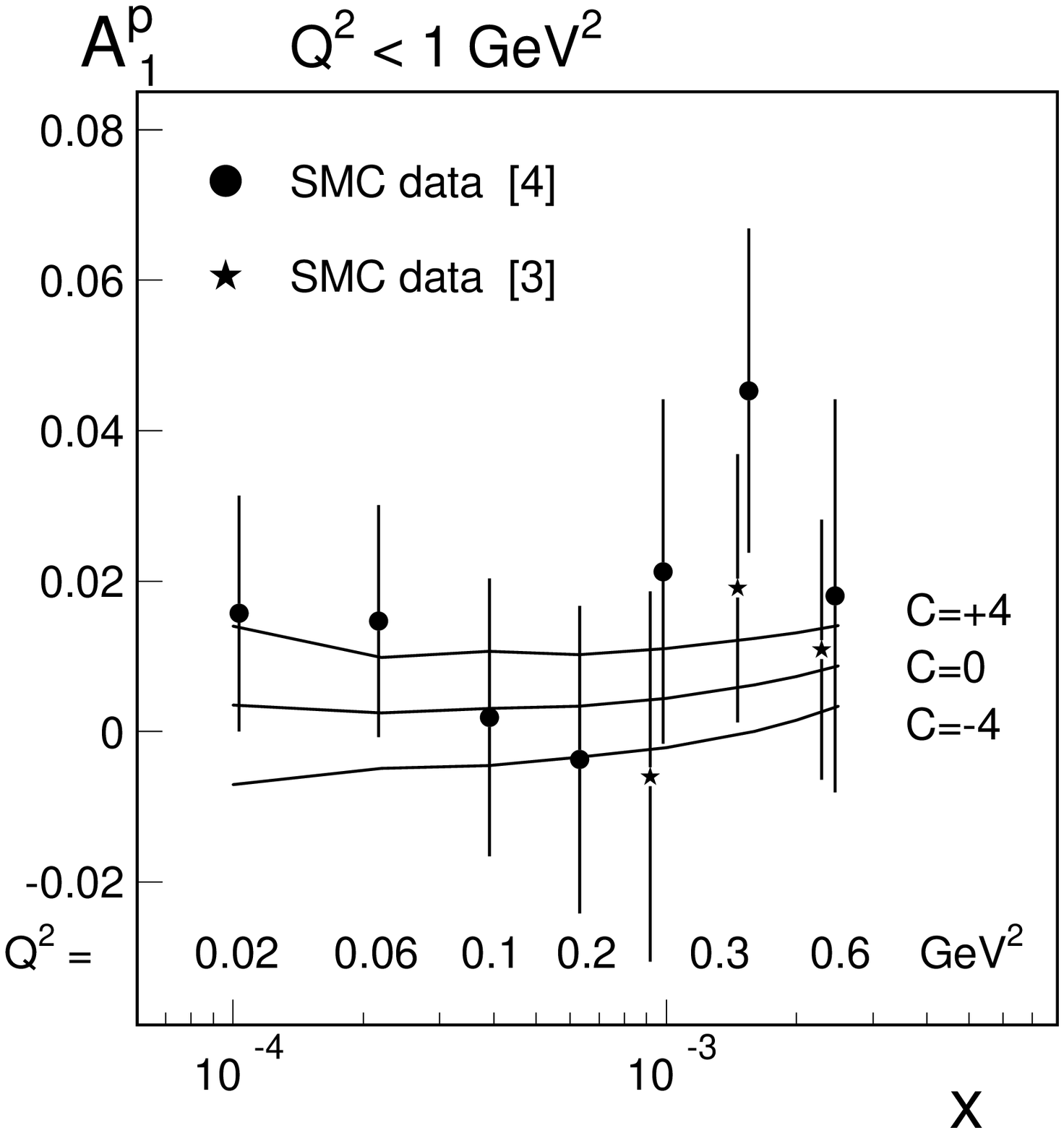}
\epsfxsize=8cm
\epsffile{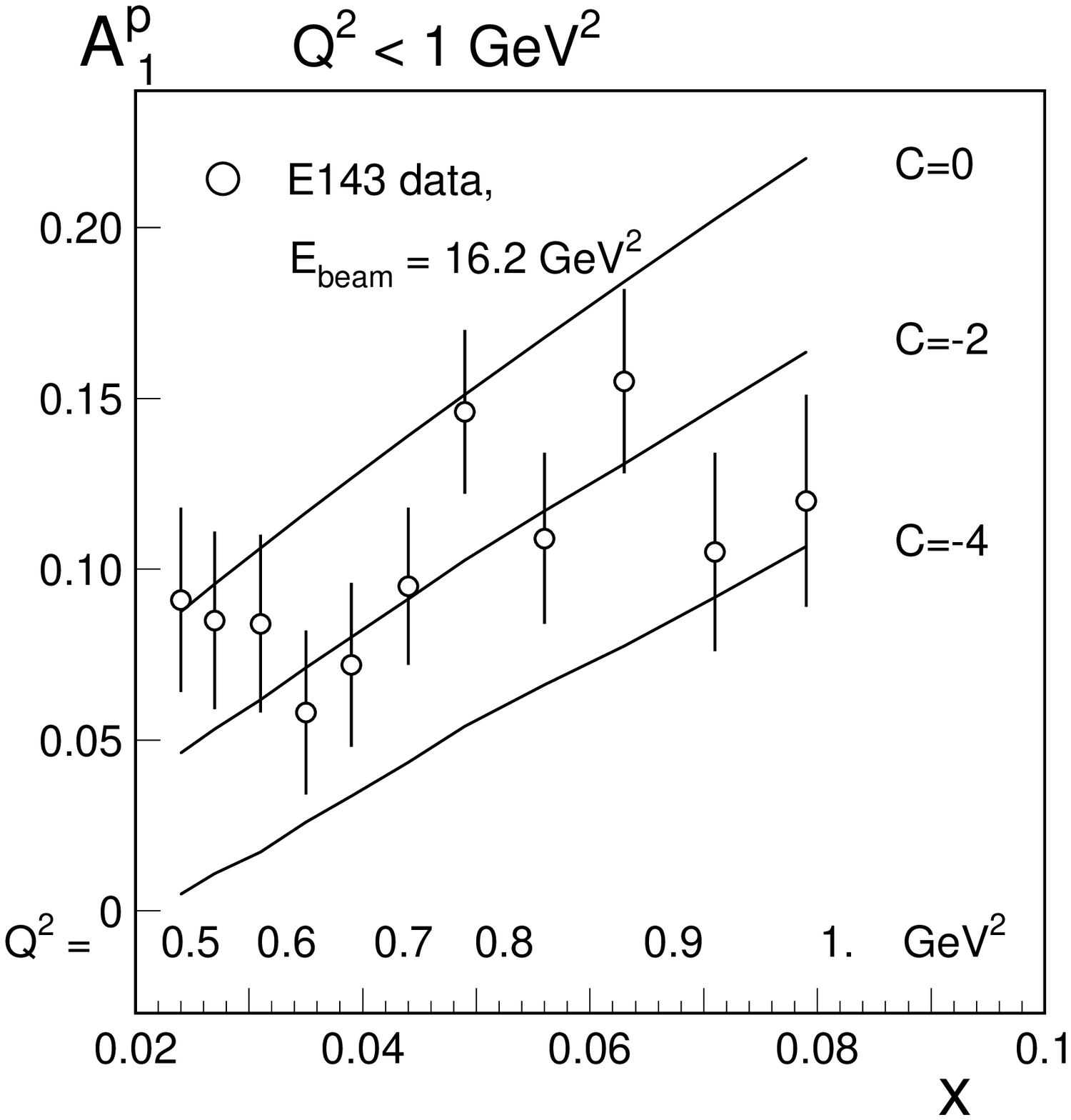}
\hfil
\caption{The asymmetry $A_1$ for the proton as a function
of $x$ at the measured $Q^2$ (marked above the $x$ axis),
obtained by the SMC \protect\cite{optimal,t15} and 
SLAC E143 \protect\cite{e143} (at 16.2 GeV incident energy). 
Errors are statistical. Curves are calculated according to Eqs
(\protect\ref{gp1}) -- (\protect\ref{vmdg1}) assuming 
different values of $C$ in Eqs (\protect\ref{nsvmd}) and (\protect\ref{svmd}).
\label{fig:a1p}}
\end{figure}

A dependence of the structure function $g_1(x,Q^2)$ given by Eqs
(\ref{gp1}) -- (\ref{g1tot}) on the parameter $C$
in Eqs (\ref{nsvmd}) and (\ref{svmd}) is illustrated in  
Fig.2 where we plot the asymmetries $A_1(x)$ for the proton 
at the measured $Q^2$ and for $Q^2<$ 1 GeV$^2$. We expect the VMD 
contribution to be dominant there. 
This cut selected the SMC data \cite{optimal,t15} at low values
of $x$ and the SLAC E143 measurements \cite{e143} at 16.2 GeV incident 
electron energy at higher $x$.\footnote{The E143 measured $A_1$ for $Q^2 <$ 
1 GeV$^2$ also at 9.7 GeV incident electron energy. For these data 4$\leq W^2\lapproxeq$
10 GeV$^2$, i.e. above the resonance region but too small for our model to be
applicable.}
To obtain predictions for the asymmetry $A_1$ rather than for $g_1$
we used the model \cite{BBJK2}
for the $F_2$ and two different parametrisations \cite{e143,t15} for $R$, 
as employed in the E143 and SMC analyses respectively. \\

The statistical accuracy of the SMC data is too poor to constraint
the value of the coefficient $C$, i.e. of the VMD--type
 nonperturbative contribution to the structure function $g_1(x,Q^2)$ at 
low values of $Q^2$. The SLAC E143 data apparently prefer a small
negative value of C. The model prediction without VMD contribution (C=0) is 
systematically higher than the E143 measurements. 
The fact that the data prefer negative 
value of the VMD contribution is consistent with the results obtained from 
the phenomenological analysis of the sum-rules \cite{IOFFE}.  \\
%There is thus a room for nonpartonic contributions to $g_1^p$.

Similar analysis performed for the neutron and deuteron structure functions,
$g_1^n$ and $g_1^d$, where in the former case data cover narrower kinematic 
interval and in the latter the statistics at low $x$ is substantially poorer, 
turned out to be inconclusive. 
   
%Predictions based on equations (\ref{dpi},\ref{gp1}) 
%shown as a curve in Fig.1 
%%reproduce a general trend in the data.

\section*{5. Summary and conclusions}
\label{section:sum}

\noindent
We have analysed the recent $g_1(x,Q^2)$ measurements at low values of $x$ and
$Q^2$ within a formalism based on unintegrated spin dependent parton 
distributions incorporating the leading order Altarelli--Parisi evolution
and the double ln$^2(1/x)$ resummation at low $x$. A VMD--type nonperturbative
part was also included since low values of $x$ in the measurements 
correlate with low values of $Q^2$. 
%at $x$=0.0001, average $Q^2$ is about 0.02 GeV$^2$. 
The ln$^2(1/x)$ effects are not yet important in the kinematic range
of the fixed target experiments but the formalism based on unintegrated
parton distributions, summarised by Eq.(\ref{dpi}),
 is very suitable for extrapolating $g_1$ to the region
of low $Q^2$.
The model reproduces a general trend in the data for the proton. 
The statistical accuracy of the SMC measurements
taken at lowest values $x$, $x > $0.00006, and of the $Q^2$,
$Q^2 >$ 0.01 GeV$^2$, is however too poor to constraint the VMD 
contribution. A more accurate data from the SLAC E143 experiment, 
where $x >$ 0.02 and $Q^2 >$ 0.5 GeV$^2$ seem to prefer a nonzero 
and negative contribution of the VMD to $g_1$ of the proton.  

\section*{Acknowledgements}
\label{section:ack}
This research was partially supported by the Polish State Committee for
Scientific Research (KBN) grants 2~P03B~132~14, 2~P03B~89~13 and by the
EU Fourth Framework Programme `Training and Mobility of Researchers', Network
`Quantum Chromodynamics and the Deep Structure of Elementary Particles',
contract FMRX--CT98--0194.  

\end{document}